\documentclass[]{article}
\usepackage{geometry}
\usepackage{graphicx}
\usepackage{subcaption}
\usepackage{amsmath}
\usepackage{amssymb}
\usepackage{enumerate}
\usepackage{framed}
\usepackage{enumitem}
\newtheorem{prop}{Prop.}
\newcommand{\ave}[1]{\langle #1 \rangle}
\newcommand{\avE}[1]{\langle #1 \rangle\!{}_{_E}}
\newcommand{\avEd}[1]{\langle #1 \rangle\!{}_{_{E+\Delta E}}}
\newcommand{\ME}{{\mathcal{M}_E}}
\newcommand{\MEe}{{\mathcal{M}_{E+\varepsilon}}}
\newcommand{\Mee}{{\mathcal{M}(E,\varepsilon)}}
\newcommand{\MEd}{{\mathcal{M}(E,\Delta E)}}
\newcommand{\lie}{\mathcal{L}}
\newcommand{\lieH}{\mathcal{L}_{X_H}}
\newcommand{\SE}{{\Sigma_E}}
\newcommand{\SEe}{{\Sigma_{E+\varepsilon}}}
\newcommand{\SEd}{{\Sigma_{E+\Delta E}}}
\newcommand{\dmu}{{\mathrm{d}\mu}}
\newcommand{\volS}{{\mathrm{Vol}(\SE)}}
\newcommand{\volSd}{{\mathrm{Vol}(\SEd)}}
\newcommand{\volM}{{\mathrm{Vol}(\ME)}}
\newcommand{\volMd}{{\mathrm{Vol}(\MEd)}}
\title{On the Generalised Equipartition Law}
\author{Guido Magnano \& Beniamino Valsesia\\ \small Department of Mathematics ``G.~Peano'', Universit\`a degli Studi di Torino, Turin, Italy}
\begin{document}
\maketitle
 
\begin{abstract}
We observe that the so-called Generalised Equipartition Law for hamiltonian systems is actually valid only under specific hypotheses -- unfortunately omitted in some textbooks -- which limit its applicability when dealing with nonlinear systems. We introduce a new coordinate--independent generalisation which overcomes this problem, and moreover can be applied to a larger set of functions. A simple example of application is discussed.
\end{abstract}

\section{Introduction}
The equipartition theorem is a celebrated result of classical statistical mechanics, and its importance could hardly be overemphasized. In an article by Tolman \cite{tolman}, dating back to 1918, one can find a generalised version  which now appears in many textbooks on statistical mechanics. The Generalised Equipartition Law states the following: 
\begin{prop}\label{due}
For any Hamiltonian $H$ depending on the $2n$ phase-space coordinates $x^i\equiv(q^\lambda,p_\lambda)$ the following equality holds:
 \begin{equation}\ave{x^i\,\frac{\partial H}{\partial x^j}}=\delta_j^i\,kT,\label{uno}\end{equation} 
 $\delta_i^j$ being the Kronecker delta ($\delta_j^i=1$ if $i=j$, $\delta_j^i=0$ if $i\ne j$).
\end{prop}
In this statement, each of the coordinates $x^i$ can be either a configuration coordinate $q^\lambda$ or a conjugate momentum $p_\lambda$. This proposition was first formulated and proven by Tolman for $i=j$; in this case, if the Hamiltonian is fully quadratic, it reproduces the classical form of the equipartition theorem. The statement was first proven for the canonical ensemble averages of functions; upon assuming that the system is ergodic, one concludes that equipartition holds for the time averages. In Kubo's book \cite{kubo}, p.~112, the full statement is proven under the condition that the potential appearing in the Hamiltonian tends to infinity when the configuration coordinates $q^\lambda$ tend to \emph{``the ends of their domain''}. 

In Kerson Huang's textbook \cite{Huang} the statement is instead proven for the microcanonical averages, and no hypothesis is made on the Hamiltonian nor on the coordinates. Indeed, from the statement of the theorem one should exclude that $x^j$ be a cyclic coordinate, i.e., that  $\frac{\partial H}{\partial x^j}\equiv 0$, for in this case $\ave{x^i\,\frac{\partial H}{\partial x^j}}\equiv 0$ even for $i=j$: this condition is so obvious, apparently, that it is generally omitted. 
The generalised equipartition law not only provides the equilibrium average for a larger set of functions (not only the additive terms of a quadratic Hamiltonian), but apparently applies to genuinely nonlinear systems; in spite of that and of Tolman's expectations, however, it did not play a central role in the development of classical statistical mechanics. In 1984, Buchdahl \cite{Buchdahl} wrote \emph{``It seems that (1) is rarely invoked to derive specific results. Indeed, I have been unable to locate any application of it other than the demonstration that the mean energy per particle of an ideal gas must exceed $\frac{3}{2}kT$ when relativistic effects are taken into account and tends to $3kT$ in the ultrarelativistic limit.''} 

However, there is an issue which seems to have gone almost unnoticed. The statement of Prop.~\ref{due}, as a whole, is coordinate-dependent; the functions $x^i\,\frac{\partial H}{\partial x^j}$ themselves have no intrinsic meaning. This would not be a problem, if the proposition were true in any coordinate system: but it is not so. 

From the proof given in \cite{Huang} it is apparent that the coordinates $x^i$ should be such that the invariant measure for the Hamiltonian system coincides with the Lebesgue measure $dx^1\wedge\ldots\wedge dx^{2n}$. This holds for any \emph{natural} chart formed by Lagrangian coordinates and conjugate momenta, $(q^\lambda,p_\lambda)$, but is actually true for any system of \emph{canonical} coordinates. Berdichevsky \cite{Berd93} introduced a generalisation to the case of arbitrary (global) non--canonical coordinates, with a suitable modification of the l.h.s.~of (\ref{uno}), showing that the only other change in this case is the appearance on the r.h.s.~of a different quantity $T'$, which coincides with the temperature $T$ if and only if the coordinate transformation is volume--preserving.

Hence, if we consider for instance the Hamiltonian of a two-dimensional harmonic oscillator, we would expect (\ref{uno}) to be true also for action-angle variables: but it is easy to see that it is not. 

In fact, let $(I_\mu,\varphi^\mu)$ be a set of action-angle coordinates, such that the Hamiltonian becomes $H=\omega_{(1)}I_1+\omega_{(2)}I_2$, the constants $\omega_{(\mu)}$ being the characteristic frequencies; then, according to Prop.~\ref{due} we should have $\ave{I_\mu\,\frac{\partial H}{\partial I_\nu}}=\delta_\mu^\nu\,kT$. But since $\frac{\partial H}{\partial I_\nu}=\omega_{(\nu)}$, one has $\ave{I_\mu\,\frac{\partial H}{\partial I_\nu}}=\omega_{(\nu)}\ave{I_\mu}$: this cannot vanish for $\mu\ne\nu$ unless $\ave{I_\mu}=0$, which would yield $\ave{I_\mu\,\frac{\partial H}{\partial I_\nu}}=0$ also for $\mu=\nu$.\footnote{The system is obviously not ergodic: in this example, the ensemble averages do not coincide with the time averages (since the action coordinates $I_\mu$ are constants of the motion, their time averages coincide with their initial values). We stress that  Prop.~\ref{due} is supposed to hold for ensemble averages, without assuming ergodicity.}

Indeed, the oscillator Hamiltonian in action-angle coordinates does not meet the requirements of Kubo's version of the generalised equipartition law (the Hamiltonian is not split into kinetic and potential terms). The statement of the law in \cite{Huang}, on the other hand, does not mention this condition, which is not used in the proof. A careful look at the proof, however, gives a clue about the origin of the apparent paradox: the operations performed on the integrals defining the ensemble averages actually require that the coordinates be globally defined, i.e., that a single coordinate system covers the whole phase space, and that all the functions involved in the proof are everywhere smooth in the integration domain. Thus, the theorem cannot be applied to action-angle coordinates, which are not global (each angular coordinate is defined only in the open interval $(0,2\pi)$, and neither action nor angle coordinates are defined around the ground state, i.e.~the minimum of the total energy). 

The fact that Prop.~\ref{due} holds only for global coordinates, however, does not rule out only canonical transformations to action-angle coordinates. As a matter of fact, it prevents the generalised equipartition law to be applicable to generic systems where the \emph{configuration space} is not diffeomorphic to $\mathbb{R}^n$. In this form, equipartition cannot be applied, for instance, to a simple Hamiltonian such as $H(q,p)=\frac{1}{2m}p^2-k\cos(q)$, the coordinate $q\in [-\pi,\pi]$ being an angle: as we shall see in the last section, in this case Prop.~\ref{due} gives \emph{completely wrong} predictions of the time averages for the function $q\frac{\partial H}{\partial q}$. 

To be clear: what we are questioning here is not the formal correctness of the proof, but the actual validity of the statement.

Most classical studies about equipartition deal with non--integrable perturbations of a harmonic oscillator, expressed in global Cartesian coordinates: for such cases, Prop.~\ref{due} is fully useful. But if one considers systems with nonlinear constraints, such as a rigid body subject to a force, then the configuration space cannot be covered by a single, global system of Lagrangian coordinates, and nothing ensures that the equipartition property holds true\footnote{Indeed, in the case of a \emph{free} rigid body the statement of Prop.~\ref{due} is valid. The reason is that all in that case all configuration coordinates are cyclic: the Hamiltonian is purely quadratic and only depends on the components of the angular momentum.}. 

Upon assuming that in natural coordinates the Hamiltonian is always quadratic in momenta, it seems that -- at least -- equipartition of \emph{kinetic} energy should be a universal property; but it has been observed \cite{Uline} that even this fails to be true if the system includes molecules of different mass and computations are done in the center--of--mass reference frame (which amounts to imposing a linear constraint on momenta).

In the last decades, violations of the equipartition law for classical hard-sphere molecular dynamics have been observed \cite{Shirts}, while other authors considered modifications of the law for the case of a confining potential \cite{Mello} and found inconsistencies related to the use of non--cartesian coordinates \cite{Rey}.

While performing numerical ergodicity tests, it is therefore important to realise that a lack of equipartition experimentally observed for time averages may not depend on a violation of the ergodicity hypothesis, but rather on the fact that Prop.~\ref{due} is already violated at the level of \emph{ensemble} averages: we shall extensively discuss in the last section a simple example. 

 Therefore, it seems quite desirable to have at our disposal an intrinsic (i.e.~coordinate--independent) statement, including Prop.~\ref{due} as a particular case. In this article we provide and prove such a statement.
\section{Intrinsic Generalised Equipartition Law}
\begin{framed}
\begin{prop}\label{tre}
Let $H:T^*Q\rightarrow\mathbb{R}$ be the Hamiltonian of an autonomous mechanical system on a configuration manifold $Q$; assume that the system has an equilibrium ground state, i.e.~that $H$ has a lower bound. Let $X$ be any (globally defined, nonsingular) vector field on the phase space, and let $X(H)$ be the derivative of $H$ along $X$. Assume that the hypersurface $H=E$ is compact for a given regular value $E$, and let $\avE{f}$ denote the (microcanonical) ensemble average of a function $f$ over the hypersurface $H=E$. Let $\dmu$ be the invariant Liouville measure on $T^*Q$, let $\ME $ be the domain in the phase space defined by $H\le E$ and let $\mathrm{Vol}(\ME)=\int_{\ME }\!\dmu$ be its volume. Then
\begin{equation}
\avE{X(H)}=\frac{k\,T}{\mathrm{Vol}(\ME)}\int_{\ME}\!\!\!\operatorname{div}(X)\,\dmu .\label{wow}
\end{equation}
\end{prop} 
\end{framed}
\par
Prop.~\ref{due} is the particular case of this statement for $X=x^i\,\frac{\partial }{\partial x^j}$. Assuming that the coordinates are canonical, $\mathrm{div}\!\left(x^i\,\frac{\partial }{\partial x^j}\right)=\delta^i_j$: hence, $\displaystyle\int_{\ME }\!\!\!\operatorname{div}\!\left(x^i\,\frac{\partial }{\partial x^j}\right)\dmu =\mathrm{Vol}(\ME )\delta^i_j$ and one recovers Prop.~\ref{due}. 

The new coordinate--independent formulation, however, goes beyond Tolman's formulation. For instance, the generalisation to non-canonical coordinates introduced by Berdichevksy \cite{Berd93} can also be obtained from (\ref{wow}): assuming that the volume form in a given coordinate system is described by a nonconstant density $\rho$, one has $\mathrm{div}(X)=\frac{1}{\rho}\frac{\partial(\rho X^\mu)}{\partial x^\mu}$. Then, applying (\ref{wow}) to $X=\Delta\, x^i\frac{\partial }{\partial x^j}$, with $\Delta=\frac{1}{\rho}$, one directly finds eqs.~(1.5) and (2.9) of \cite{Berd93}\footnote{it is apparent in this way that the quantity $T'$ in \cite{Berd93}, which coincides with $T$ if $\rho\equiv 1$, depends on the chosen coordinate system and therefore has no intrinsic thermodynamical interpretation.}. 

Eq.~(\ref{wow}) might be obtained through a procedure which is reminiscent of the proof of Prop.~\ref{due} in \cite{Huang}:
\begin{align*}
    \bigg\langle X^\mu\frac{\partial H}{\partial x^\mu} \bigg\rangle
    &=\frac{1}{\volS}\lim_{\epsilon\rightarrow 0}\frac{1}{\epsilon}\bigg[\int_{\MEe}X^\mu\frac{\partial H}{\partial x^\mu}\dmu-\int_{\ME}X^\mu\frac{\partial H}{\partial x^\mu}\dmu\bigg]
    \\
    &=\frac{1}{\volS}\frac{\partial}{\partial E}\int_{\ME}X^\mu\frac{\partial H}{\partial x^\mu}\dmu
    =\frac{1}{\volS}\frac{\partial}{\partial E}\int_{\ME}X^\mu\frac{\partial (H-E)}{\partial x^\mu}\dmu;
\end{align*}
next, one applies Stokes' theorem to the last integral. The function $ (H-E)$ obviously vanishes on the boundary hypersurface $H=E$, so the boundary integral cancels out and one finds 
\begin{align*}
    & \frac{1}{\volS}\frac{\partial}{\partial E}\int_{\ME}X^\mu\frac{\partial (H-E)}{\partial x^\mu}\dmu =-\frac{1}{\volS}\frac{\partial}{\partial E}\int_{\ME}(H-E)\frac{\partial X^\mu}{\partial x^\mu}\dmu
    \\
    & =\frac{1}{\volS}\int_{\ME}\frac{\partial X^\mu}{\partial x^\mu}\dmu=\frac{k T}{\volM}\int_{\ME}\mathrm{div}(X)\dmu.
\end{align*}
(while taking the derivative w.r.~to $E$ one should indeed consider the dependence on $E$ of the integration domain $\ME$, but the resulting term cancels out, again, because $H=E$ on the boundary of $\ME$; the fact that $\frac{1}{\volS}=\frac{k T}{\volM}$, instead, is a consequence of the microcanonical definition of temperature and will be discussed below). However, this derivation requires that a single coordinate system covers all the integration domain $\ME$ (otherwise, Stokes' theorem cannot be applied in this way), which is exactly the crucial limitation that we seek to overcome. 

But the statement of the generalized equipartition law in Prop.~\ref{tre} is now coordinate-independent, and therefore one can attempt to find a coordinate-free proof: this will be done in the next section. The proof requires differential-geometric methods, and for this purpose we shall first introduce a suitable construction of the microcanonical measure, connected with the symplectic structure of the phase space. 

Prop.~\ref{tre} shows that to assess equipartition the existence of a global coordinate system is not necessary. The crucial condition, instead, concerns the vector field $X$: if $X$ has singular points in $\ME$, then the hypersurface integral on the l.h.s.~of eq.~(\ref{wow}) may be different from zero even if $X$ is divergenceless almost everywhere\footnote{This is strictly analogous to a well-known situation in electrostatics. Consider a point charge located at $\boldsymbol{x}$: the vector field $\vec{E}$ generated by the point charge is singular at $\boldsymbol{x}$ and its flux through a closed surface surrounding $\boldsymbol{x}$ does not vanish, although $\vec{E}$ is divergenceless at any other point.}. This explains why Prop.~\ref{due} is violated if one takes action--angle coordinates: the field $I_\lambda\,\frac{\partial }{\partial I_\mu}$ extends to a globally defined vector field only if $\lambda=\mu$, while for $\lambda\ne\mu$ it becomes singular for $I_\lambda\rightarrow 0$.

\section{A geometrical microcanonical measure}
In order to prove Prop.~\ref{tre}, we need an intrinsic definition of the microcanonical measure. The latter is usually defined as a limit of the Liouville invariant measure on the domain $E\le H\le (E+\varepsilon)$ when $\varepsilon\rightarrow 0$ (see e.g.~\cite{Huang}). Let $\SE$ be the energy hypersurface (i.e.~the level set $H=E$), and let $dq^1\wedge dq^2\wedge\ldots\wedge dp_n$ the Lebesgue measure on $T^*Q$. One defines the total volume of  $\SE$  by 
\begin{equation}\mathrm{Vol}(\SE)= \lim_{\varepsilon\to 0}\frac{1}{\varepsilon }\left[\int_{\MEe}\!\!\!\!\! dq^1\wedge\ldots\wedge dp_n-\int_{\ME }\!\!\!dq^1\wedge\ldots\wedge dp_n\right];\label{microv}\end{equation} 
 the microcanonical average (at total energy $E$) of any $L^1$ function $F$ defined in a neighbourhood of $\SE$ is then defined as in~\cite{Huang,Chand}: 

\begin{equation}
\avE{F}= \frac{1}{\mathrm{Vol}(\SE)}\lim_{\varepsilon\to 0}\frac{1}{\varepsilon}\bigg[\int_{\MEe}\!\!\!\!\! F\,dq^1\wedge\ldots\wedge dp_n-\int_{\ME }\!\!\! F\,dq^1\wedge\ldots\wedge dp_n\bigg].\label{micro}
\end{equation}

In alternative, the microcanonical measure can be introduced as a suitable Dirac distribution relative to the energy hypersurface \cite{Chand, Buonsante_2016}. Both definitions, however, are impractical if no global coordinate system is available: each integral can be defined only within the domain of a coordinate chart, and to extend integrals to $\ME$ one should rely, in principle, on a partition of unity adapted to the chart domains. In the sequel we follow a different, more geometric approach.

Whenever $E$ is a \emph{regular value} for the Hamiltonian $H$ (i.e.~if $dH$ does not vanish at any point of $\SE$), $\SE$ is a smooth manifold of dimension $2n-1$, and (if $H$ is lower bounded) it coincides with the boundary of the domain $\ME$:\mbox{ $\SE\equiv\partial\ME$.} Since the Hamiltonian is conserved, the manifold $\SE$ is invariant under the Hamiltonian flow.

To integrate functions over $\SE$, what we need is a \mbox{$(2n-1)$--form} nowhere vanishing on $\SE$; to be identified with the microcanonical measure, up to overall normalisation, this form has to be invariant under the Hamiltonian flow.

Notice that there is a standard procedure to define the restriction of a volume form to a submanifold, if the ambient manifold is endowed with a Riemannian metric. Now, the phase space of a Hamiltonian system (a cotangent bundle, in the setup of classical mechanics) is always endowed with an invariant volume form, but there is no natural Riemannian structure. Indeed, any phase space being a differentiable manifold can be endowed with infinitely many Riemannian structures: but each of them would produce a different measure on the submanifold $\SE$, and we would need to single out those that are invariant under the Hamiltonian flow. We shall instead adopt a different strategy. 

We start from the natural volume form on $T^*Q$, which is (up to a constant factor) the \emph{$n$--th exterior power} of the canonical symplectic form \mbox{$\omega = dp_i\wedge dq^i$.} For any system of canonical coordinates, it coincides with the Lebesgue measure $dq^1\wedge dq^2\wedge\ldots\wedge dp_{n-1}\wedge dp_n \equiv  \frac{1}{n!}\,\omega^n$. 

In the sequel, we shall denote by $\dmu$ this volume $n$--form. This notation is closer to the measure--theoretic usage than to the differential--geometric setup that we adopt here, resulting in a somehow hybrid notation, but we feel that for most readers the formulae will be clearer if we write $\dmu$ instead of $\frac{1}{n!}\,\omega^n$ (we stress that in the latter expression the denominator $n!$ is merely due to the definition of wedge product: $n$ is the number of degrees of freedom -- i.e.~half the dimension of the phase space). 

Our aim is decomposing $\dmu$, in a neighbourhood of $\SE$, into the wedge product of a $1$--form and a \mbox{$(2n-1)$--form,} in such a way that the latter defines a volume form on $\SE$ and is invariant along the hamiltonian flow generated by $H$. 

We shall need the following notations and properties. We denote by $i_{X}\theta$ the \emph{interior product} of a differential  $p$--form $\theta$ and a vector field $X$. If $\theta$ is a 1--form, then $i_{X}\theta$ is nothing but the evaluation of $\theta$ on $X$, i.e.~$\langle\theta,X\rangle$. For any function $f$ the interior product with a vector field is always zero, $i_Xf\equiv 0$, which entails that $i_X(f\theta)=fi_X\theta$ for any $p$--form $\theta$. 

We shall denote by $\lie_X\theta$ the \emph{Lie derivative} of $\theta$ along $X$:  \mbox{$\lie_X\theta=d\left(i_X\theta\right)+i_X(d\theta)$} (Cartan formula)\cite{abrahm}. The condition \mbox{$\lie_X\theta= 0$} ensures that $\theta$ is invariant under the flow generated by $X$. For any Hamiltonian vector field $X_H$ the symplectic form is conserved, $\lie_{X_H}\omega=0$, and therefore the volume form $\dmu$ is invariant as well (Liouville theorem). 

Our geometrical setup is provided by the following statement:

\vbox{
\begin{framed}
\begin{prop}\label{quattro}
Let $H:T^*Q\xrightarrow{}\mathbb{R}$ be a Hamiltonian, bounded from below and such that the set of stationary points of H is discrete; let  $E$ be a regular value for $H$ such that $\ME \equiv\big\{x \in T^*Q : H(x)\le E\big\}$ is compact, $\SE\equiv\left\lbrace x \in T^*Q : H(x)=E\right\rbrace$ coincides with $\partial\ME $ and is also compact. 
Let $\alpha$ be a 1--form, defined on some neighbourhood of $\SE$ in $T^*Q$, such that
\begin{itemize}[noitemsep]
    \item $d\alpha=0$,
    \item $i_{X_H}\alpha=\langle\alpha,X_H\rangle=1$,
\end{itemize}
and define 
$
\Omega=  \frac{1}{(n-1)!}\, \alpha\wedge \omega^{n-1}.
$
Then
\begin{enumerate}[noitemsep]
\item the $(2n-1)$--form $\Omega$ is invariant under the flow generated by $X_H$;
\item $dH\wedge\Omega=\frac{1}{n!}\,\omega^n=\dmu$ on the domain where $\Omega$ is defined;
\item $\Omega$ is a volume form on $\SE$; 
\item $\volS=\displaystyle\int_{\SE}\!\!\Omega$;
\item  for any smooth function $F$ defined in a neighbourhood of  $\SE$ the mean value
$$\avE{F}=\frac{1}{\volS}\int_{\SE}\! F\,\Omega $$
coincides with the microcanonical average (\ref{micro}), and is therefore independent of the particular 1--form $\alpha$ chosen to define $\Omega$.
\end{enumerate}
\end{prop}
\end{framed}}

\subsection{Proof of Prop.~\ref{quattro}}
The assumptions on $\alpha$ imply that $\lieH\alpha=d(i_{X_H}\alpha)+i_{X_H}d\alpha=0$ \cite{abrahm};\\ thus \mbox{$\lieH\Omega=\frac{1}{(n-1)!}\left((\lieH\alpha)\wedge\omega^{n-1}+\alpha\wedge\lieH\omega^{n-1}\right)=0$,} which proves \emph{(i)}.\\ Furthermore, we recall that $d\omega=0$ and therefore $d(\omega^{n-1})=0$ as well:\\ then, \mbox{$d\Omega=\frac{1}{(n-1)!}(d\alpha\wedge\omega^{n-1}-\alpha\wedge d\omega^{n-1})=0$.} Hence, $dH\wedge  \Omega=d(H\Omega)$.\\
Since $d\alpha=0$, in some neighborhood $\mathcal{U}$ of any point of the domain of definition of $\Omega$ one can write $\alpha=df$ for some function $f:\mathcal(U)\rightarrow\mathbb{R}$ whose Poisson bracket with $H$ is $\lbrace H,f\rbrace=1$. Then, locally $d(H\Omega) =\frac{1}{(n-1)!}d(H\alpha\wedge \omega^{n-1})=\frac{1}{(n-1)!}d(H\, df\wedge \omega^{n-1})$. For any function $f$ and for the corresponding hamiltonian vector field $X_f$ one has $i_{X_f}\omega=-df$; moreover, for any vector field $X$ one has $i_X\omega^n=n(i_X\omega)\wedge\omega^{n-1}$.  Therefore, $df\wedge\omega^{n-1}=-\frac{1}{n}\,i_{X_f}\omega^n$ and
\begin{align*}
dH\wedge  \Omega & =- \frac{1}{n!}d\left(H i_{X_f}\omega^{n}\right)=-\frac{1}{n!} d(i_{X_f}(H\omega^{n}))=\\
& =-\frac{1}{n!}\left(\mathcal{L}_{X_f}(H\omega^n)-
i_{X_f}\underbrace{d(H\omega^{n})}_{=0}\right)=-\frac{1}{n!}\left(\mathcal{L}_{X_f}(H)\omega^n+H\underbrace{\mathcal{L}_{X_f}\omega^n}_{=0}\right)=\\&=-\frac{\big\{f,H\big\}}{n!}\,\omega^n=
\frac{1}{n!}\,\omega^n=\dmu. 
\end{align*}
\noindent At any point of $\SE$, let $\lbrace X_1,\ldots X_{2n-1}\rbrace$ be any set of linearly independent vectors tangent to $\SE$, i.e.~such that $i_{X_k}dH=0$, and let $Y$ be any vector which is not tangent to $\SE$ (and is therefore linearly independent of the set $\lbrace X_k\rbrace$). By \emph{(ii)}, that we have just proven, one has $\omega^n(Y,X_1,\ldots X_{2n-1})= n!~ i_{Y}dH\cdot\Omega(X_1,\ldots X_{2n-1})$, so the latter cannot vanish and $\Omega$ is thus a good volume form on $\SE$.\\
To prove \emph{(iv)} and \emph{(v)}, we first observe that \emph{(iii)} and $d\Omega=0$ imply that
$$
\int_{\ME }\!\!\!\! F\dmu=\int_{\ME}\!\!\!\! d(FH\Omega)-\int_{\ME }\!\!\!\! H dF\wedge\Omega.
$$
By Stokes' theorem, $\displaystyle\int_{\ME}\!\!\!\! d(FH\Omega)=\displaystyle\int_{\SE}\!\!\!\! FH\Omega$; since on $\SE$ the Hamiltonian is constant, $H=E$, this integral equals $E\!\displaystyle\int_{\SE}\!\!\!\!  F\,\Omega$. Therefore,
$$
\int_{\ME }\!\!\!\! F\dmu=E\int_{\SE}\!\!\!\!  F\,\Omega-\int_{\ME }\!\!\!\! H dF\wedge\Omega.
$$
Now, let $\Mee$ be the region defined by $E\le H\le(E+\varepsilon)$. For $\varepsilon>0$ small enough, $\Mee$ is compact and is contained in the domain of definition of $\Omega$; moreover $\partial\Mee=\SEe\bigcup\SE$ (with opposite orientations). \\
To obtain the  microcanonical average (\ref{micro}) we observe that the integral of a smooth function $F$ in the region $\Mee$ is equal to
$$
\int_{\MEe }\!\!\!\!\!\!\! F\dmu-\int_{\ME }\!\!\!\!\! F\dmu=(E+\varepsilon)\!\!\int_{\SEe}\!\!\!\!\!\! F\Omega-E\!\!\int_{\SE}\!\!\! F\Omega-\int_{\MEe}\!\!\!\!\!\!\!\! H dF\wedge\Omega+\int_{\ME }\!\!\! H dF\wedge\Omega.
$$
Using again Stokes' theorem, 
$$
E\left(\int_{\SEe}\!\!\!\!\!\! F\Omega-\int_{\SE}\!\! F\Omega\right)=E\!\int_{\Mee}\!\!\!\!\!\! d(F\Omega)=E\!\int_{\Mee}\!\!\!\!\!\! dF\wedge\Omega
$$
and therefore 
$$
\lim_{\varepsilon\to 0}\frac{1}{\varepsilon}\left(\int_{\Mee }\!\!\!\!\!\!\! F\dmu\right)=\int_{\SE}\!\!\! F\Omega-\lim_{\varepsilon\to 0}\frac{1}{\varepsilon}\int_{\Mee}\!\!\!\!\!(E-H) dF\wedge\Omega.
$$
In particular, if we take $F\equiv 1$ the last integral vanishes and we find that $\volS$ -- as defined by eq.~(\ref{microv}) -- equals $\displaystyle\int_{\SE}\!\!\Omega$. \\
Finally, for arbitrary $F$, let $G$ be the function on $\Mee$ such that $dF\wedge\Omega=G\dmu$. Since $F$ is smooth and $\Mee$ is compact, $G$ has a maximum, that we denote by $g$.  Since on $\Mee$ one has $|E-H|\le\varepsilon$, 
$$\int_{\Mee}\!\!\!(E-H) dF\wedge\Omega\le \left|\int_{\Mee}\!(E-H) dF\wedge\Omega\,\right| \le \varepsilon  \left|\,g\int_{\Mee}\!\dmu\, \right|,
$$
therefore $\displaystyle\lim_{\varepsilon\to 0}\dfrac{1}{\varepsilon}\displaystyle\int_{\Mee}\!(E-H)\, dF\wedge\Omega=0$. This completes the proof.
\subsection{Existence of the 1-form $\alpha$}
Although the 1-form $\alpha$ of Prop.~\ref{quattro} does not appear in the statement of Prop.~\ref{tre}, our proof of the latter rests on  Prop.~\ref{quattro}: this raises the problem of the actual existence of a 1-form $\alpha$ with the required properties. It is evident that such a form cannot exist at stationary points of the Hamiltonian $H$, because the vector field $X_H$ vanishes at these points and the condition $i_{X_H}\alpha=1$ cannot be fulfilled. \\
Actually, we only need that $\alpha$ be defined on a neighbourhood of $\SE$; we required that $E$ be a regular value for $H$, which means that there are no points in $\SE$ where $dH=0$. Thus, it is easy to see that 1-forms with the required properties exists locally on $\SE$. In fact, one can invoke the flow--box theorem to ensure that local coordinate systems $\lbrace x^\lambda\rbrace$ exists such that  $X_H=\dfrac{\partial}{\partial x^1}$. Then, the differential $dx^1$ has the required properties. However, to make use of the Stokes theorem in the proof of Prop.~\ref{quattro} we needed that the 1-form $\alpha$ be defined on a whole neighbourhood of $\SE$, and this cannot be ensured by the flow-box theorem. Indeed, if $\SE$ is compact (as is required) the coordinate $x^1$ cannot be extended to all of $\SE$, yet there are cases where its differential $dx^1$ is globally defined on $\SE$: this is the case, for instance, of systems with one degree of freedom. \\
On the other hand, one can endow the phase space with an (arbitrary) Riemannian scalar product $(\,,\,)$ and produce the vector field $\tilde{X}=(X_H,X_H)^{-1}X_H$; this can be done globally  except at points where $X_H=0$. Then, consider the 1-form $\tilde{X}^\flat$, defined as usual by $\langle\tilde{X}^\flat,Y\rangle=(\tilde{X},Y)$ for any vector $Y$. By construction, $i_{X_H}\tilde{X}^\flat=1$. Hence, 1-forms with the latter property do exist globally in the complement of the set of stationary points of $H$: but in general they will not be closed. The property $i_{X_H}\alpha=1$ is preserved if one adds to $\alpha$ any 1-form $\beta$ such that $i_{X_H}\beta=0$: this leaves open the possibility that such a $\beta$ can be found so that the sum $\tilde{X}^\flat+\beta$ is closed.\\
In the case of integrable Hamiltonians, the $(2n-1)$--dimensional energy hypersurface $\SE$ is foliated by $n$--dimensional Arnol'd-Liouville tori. In a neighbourhood of each torus, action-angle coordinates can be defined: and even if the angular coordinates $\varphi^i$ have a discontinuity, their differentials $d\varphi^i$ are globally defined on that neighbourhood, and at least in some cases they extend to a neighbourhood of $\SE$. For each angle coordinate, one has $i_{X_H}\varphi^i=\omega_{(i)}$, where the constant $\omega_{(i)}$ is the corresponding characteristic frequency. Hence, if we take any set of constant coefficients $c_i$ such that $\sum_{i=1}^n c_i \omega_{(i)}=1$, the 1-form $\alpha=\sum_{i=1}^n c_i d\varphi^i$ has the required properties. One can check by direct calculation that the form $\Omega$ so obtained does not depend on the particular choice of the coefficients $c_i$.\\
At the moment we do not know more general conditions for the global existence of $\alpha$, therefore Prop.~\ref{tre} will be proved upon the additional assumption that $\alpha$ exists.
\subsection{Gibbs entropy and temperature}
There is a long-standing debate on whether the correct definition of entropy for the microcanonical ensemble should be the one introduced by Boltzmann, \mbox{$S=k\log\omega(E)$} (where $\omega(E)$ is the density of states with energy $E\le H \le E+\varepsilon$), or rather  the Gibbs entropy $S=k\log\volM$. The two definitions are known to be equivalent in the thermodynamic limit; we shall not enter into the debate \cite{Uline,Buonsante_2016}, but in our setup it is more natural to adopt Gibbs' definition. We have already proven that 
$$
\frac{d\volM }{dE}=\lim_{\varepsilon\to 0}\frac{1}{\varepsilon}\left(\int_{\MEe }\!\!\!\! \dmu-\int_{\ME }\!\! \dmu\right)=\volS.
$$
The temperature being given by $\dfrac{dS}{dE}=\dfrac{1}{T}$, from Gibbs' definition of the entropy $S$ we get
\begin{equation}\frac{1}{kT}=\frac{1}{\volM }\frac{d\volM }{dE}=\frac{\volS}{\volM }.\label{kT}
\end{equation}
\subsection{Proof of Prop.~\ref{tre}}
Let $X$ be a vector field without singularities in $\ME$; for $X(H)\equiv i_X dH$ we find
\begin{align*}
\avE{X(H)}&=\frac{1}{\volS}\int_{\SE}\!\left(i_X dH\right)\Omega=\\&=\frac{1}{\volS}\int_{\SE}\! i_X(dH\wedge \Omega)-\frac{1}{\volS}\int_{\SE}dH\wedge i_X\Omega.
\end{align*}
Consider now the identity $dH\wedge i_X\Omega=d(H\,i_X\Omega)-H\,d(i_X\Omega)$; by Stokes' theorem, $\displaystyle\int_{\SE}d(H\,i_X\Omega)$ should vanish because $\SE$ is the boundary of $\ME$ and therefore has no boundary, $\partial\SE=\emptyset$; in turn, $\displaystyle\int_{\SE}\!\!\! H\,d(i_X\Omega)=E\!\displaystyle\int_{\SE}\!\!\! d(i_X\Omega)$ which also vanishes for the same reason. Finally, using $dH\wedge \Omega=\dmu$, once again Stokes' theorem, the definition of divergence of a vector field $d(i_X\dmu) = \operatorname{div}(X)\dmu$  and eq.~(\ref{kT}), we find 
\begin{equation*}
\avE{X(H)}=\frac{1}{\volS}\int_{\SE}\!\! i_X\dmu=\frac{1}{\volS}\int_{\ME}\!\!\! d(i_X\dmu)=\frac{kT}{\volM}\int_{\ME }\!\!\! \operatorname{div}(X)\dmu.
\end{equation*}

\section{Beyond the harmonic oscillator}
To show to which extent the previous considerations provide new insight into equipartition anomalies, we now discuss their application to an elementary case, which -- in spite of being a system with only one degree of freedom, and quite a familiar one -- already exhibits a number of unexpected features. 

Our example will be nothing but a simple ideal  pendulum in a fixed vertical plane (a detailed description of this system from the thermodynamical viewpoint can be found in \cite{Berd97}). In the numerical computations we assumed $m$ = 1\,kg and that the pendulum length is 1\,m, using the value 9.81\,m/s$^2$ for the gravitational acceleration constant $g$. We indicate the position of the pendulum by the angle $q\in (-\pi,\pi)$, where $q=0$ corresponds to the lower equilibrium position; numerical values of the total energy $E$ are in joule. The phase space is a cylinder, and the Hamiltonian is
$$
H=\frac{p^2}{2}-g\cos(q)
$$
The minimum of this Hamiltonian is $H(0,0)=-g$; there is another critical value, $H=g$. \\ For $-g<E<g$, the motion is oscillatory and the level set $\SE$ is a closed curve surrounding the equilibrium point $(0,0)$; for $E=g$ the level set is a singular eight-shaped curve known as \emph{separatrix}, while for $E>g$ the level set $\SE$ is the union of two closed curves surrounding the cylinder (Fig.1). The system is ergodic on $\SE$ for $E<g$, while for $E>g$ it is ergodic on each connected component of $\SE$.

Numerical computation of the four time averages $\avE{f_{11}}=\avE{q\frac{\partial H}{\partial q}}$, $\avE{f_{12}}=\avE{q\frac{\partial H}{\partial p}}$, $\avE{f_{21}}=\avE{p\frac{\partial H}{\partial q}}$ and $\avE{f_{22}}=\avE{p\frac{\partial H}{\partial p}}$ shows that for $E<g$ Prop.~\ref{due} gives an exact prediction: for each $E$, the values of $\avE{f_{11}}$ and $\avE{f_{22 }}$ coincide, while the values of $\avE{f_{12}}$ and $\avE{f_{21}}$ both vanish (up to the numerical error). 

Notice that $f_{22}$ is twice the kinetic energy; for a harmonic oscillator, $f_{11}$ would be twice the potential energy, and in that case \mbox{$\avE{f_{11}}=\avE{f_{22}}$} would mean that the average kinetic energy equals the average potential energy. For the pendulum this is no longer true (in agreement with the virial theorem). 

Below the critical energy, the only fact that may be surprising is that $\avE{f_{11}}$ and $\avE{f_{22}}$ grow with $E$ up to $E\approx 7.4$, then decrease (Fig.2). Since both values should be equal to $kT$, there is a range of energies where the heat capacity of the system is negative (as already noticed in \cite{Berd97}). This seemingly unphysical situation has been observed for other systems \cite{Thirring}. It has been argued that this behaviour -- yielding a thermodynamical instability which poses some problems in astrophysics -- should disappear in the thermodynamic limit; but with a single degree of freedom we are evidently very far from that limit.  

In contrast, if one computes the time averages $\avE{f_{11}}$ and $\avE{f_{22 }}$ for energies above the critical value, $E>g$, one finds that $\avE{f_{22}}$ increases monotonically with $E$, while $\avE{f_{11}}$ is always lower: it attains a maximum at approx.~$E=14.2$, then starts decreasing and slowly tends to a constant value for $E\rightarrow\infty$ (Fig. 3). 

Hence, Prop.~\ref{due} completely fails to predict the values of $\avE{f_{11}}=\avE{q\frac{\partial H}{\partial q}}$ for $E>g$, although natural coordinates in the phase space are used.

The numerical observations are instead correctly predicted if we use our approach. In fact, $f_{22}$ is the derivative of $H$ with respect to the vector field $p\frac {\partial}{\partial p}$, which is globally defined on the phase space; thus Prop.~\ref{tre} applies, and eq.~(\ref{wow}) gives the correct result for any energy. 

For a function such as $\frac{1}{3}\,p^4\sin(q)^2$, which is the derivative of $H$ along the vector field $X=\frac{1}{3}\,p^3\sin(q)^2\frac{\partial}{\partial p}$, the time averages cannot be obtained from Prop.~\ref{due}, while the value given by eq.~(\ref{wow}) is in full agreement with the time average that we have obtained by numerical simulation, for different energies, even if the divergence $\operatorname{div}(X)=p^2\sin(q)^2$ is not constant.\footnote{This function has no particular significance: we have just chosen a function which extends smoothly to the whole phase space, has nonvanishing ensemble average and is obtained by deriving the Hamiltonian along a vectorfield with nonconstant divergence.} 

If, instead, we consider the field $q\frac{\partial}{\partial q}$ which defines the function $f_{11}$, we see that it is discontinuous for $q\rightarrow\pm\pi$. As long as $E<g$, the domain $\ME$ does not intersect the line of discontinuity, so eq.~(\ref{wow}) still holds true. When $E>g$, the field $X$ is discontinuous on $\ME$, so Prop.~\ref{tre} does not apply, despite the fact that $f_{11}$, by itself, is everywhere well defined. This explains why for $f_{11}$ the usual equipartition formula ceases to work exactly when the critical energy is surpassed. 

As a matter of fact, our geometrical setup does allow one to predict the exact behaviour of the time average of $f_{11}$ for $E>g$. Let us give a closer look to the objects involved. For this system, $n=1$ and the volume form $\Omega$ on $\SE$ is completely defined by the two requirements $dH\wedge d\Omega=dp\wedge dq$ and $d\Omega=0$. The first requirement is equivalent to $i_{X_H}\Omega=1$; the form $\Omega$, for $n=1$, coincides with the 1-form $\alpha$ in Prop.~\ref{quattro}. For any point $(q,p)$ in the phase space, the elapsed time from the configuration $(0,\sqrt{H(q,p)+g})$, which lies on the same orbit, is given by
$$
T(q,p)=\int_{0}^{q}\frac{ds}{\sqrt{2H(q,p)+2g\cos(s)}}
$$ 
Although the function $T$ is defined only for $q\in(-\pi,\pi)$, its differential $dT$ extends to the whole phase space except for the two stationary points where $dH=0$. It is easy to see that $X_H(T)=1$, so we can set $\Omega=dT$. The microcanonical measure of any arc of $\SE$ defined in this way is nothing but the time of permanence, so the ensemble average of any function is automatically identical to the time average. The volume $\volS$ is nothing else than the orbit period. Under these premises, Prop.~\ref{quattro} tells us that for the function $f_{11}$ one has 
$$
\avE{f_{11}}=\frac{1}{\volS}\int_{\SE}\! f_{11}\,\Omega.
$$
The problem arises with the subsequent step needed to recover integration over $\ME$. It is still true that 
$$
\avE{f_{11}}=\frac{1}{\volS}\int_{\SE}\!\left(i_X dH\right)\Omega=\frac{1}{\volS}\int_{\SE}\! i_X(dH\wedge \Omega)=\frac{1}{\volS}\int_{\SE}\! i_X\dmu,
$$
because $\int_{\SE}(i_X\Omega) dH$ vanishes. But  now we cannot apply Stokes' theorem to the domain $\ME$, bounded by $\SE$, because $X$ is discontinuous on the line $q=\pm\pi$. However, let us take a reference energy $E>g$ and a positive energy difference $\Delta E$: assuming that $p>0$ along the orbit (i.e., the pendulum is rotating counterclockwise), we can consider the line $\Gamma$ which is formed (Fig.~4) by 
\begin{itemize}
\item the arc of the curve $\SEd$ with $p>0, q\in[-\pi+\varepsilon,\pi-\varepsilon]$, with positive orientation, 
\item the arc of the curve $\SE$ with $p>0, q\in[-\pi+\varepsilon,\pi-\varepsilon]$, with negative orientation, 
\item the segment $\gamma^+$ of the line $q=-\pi+\varepsilon$ connecting $\SE$ to $\SEd$,
\item  the segment $\gamma^-$ of the line $q=\pi-\varepsilon$ connecting $\SEd$ to $\SE$.
\end{itemize}
In the limit $\varepsilon\rightarrow 0$ the curve $\Gamma$ is the boundary of a region that coincides with the $p>0$ component of $\MEd$, and we have
$$
\frac{1}{2}\Big(\volSd\avEd{f_{11}}-\volS\avE{f_{11}}\Big)=\int_{\Gamma}\! i_X\dmu-\int_{\gamma^+}\! i_X\dmu-\int_{\gamma^-}\! i_X\dmu,
$$
where the factor 1/2 on the l.h.s.~is due to the fact that we are integrating only on one half of each energy level set. \\
Now, the vector field $X$ is smooth in the domain bounded by $\Gamma$, therefore we can apply Stokes' theorem to the first integral. If $X$ were not discontinuous, the two integrals on $\gamma^+$ and $\gamma^-$ would cancel each other and we would obtain the same result as produced by eq.~\ref{wow}. Here, instead, for $X=q\frac{\partial}{\partial q}$ one has $i_X\dmu=-qdp$: the integrand on $\gamma^+$ is thus $-\pi dp$, while on $\gamma^-$ the integrand is $\pi dp$. The two integrals have opposite orientation, so they sum up to give a total contribution of $2\pi\left(\sqrt{2(E+\Delta E-g)}-\sqrt{2(E-g)}\right)=2\pi\Delta p$. Hence we obtain
 $$
 \frac{1}{2}\Big(\volSd\avEd{f_{11}}-\volS\avE{f_{11}}\Big)=\frac{1}{2}\volMd-2\pi\Delta p
$$
This gives a precise description of the variation of the time averages of $f_{11}$ above the critical energy; in particular, for $E\rightarrow\infty$ the effect of gravity becomes negligible and the orbits in the phase space tend to circles with constant $p$: the area of the region between two such circles being exactly $2\pi\Delta p$, the r.h.s.~of the formula above tends to zero. As for the l.h.s., the orbital period tends to zero for energy $E\rightarrow\infty$, and for fixed $\Delta E$ the ratio $\volSd/\volS$ tends to 1. This explains why the time average $\avE{f_{11}}$ tends to a constant.

\begin{figure}[h]
\centering
\begin{subfigure}[t]{0.55\textwidth}
\centering
\includegraphics[width=0.9\linewidth]{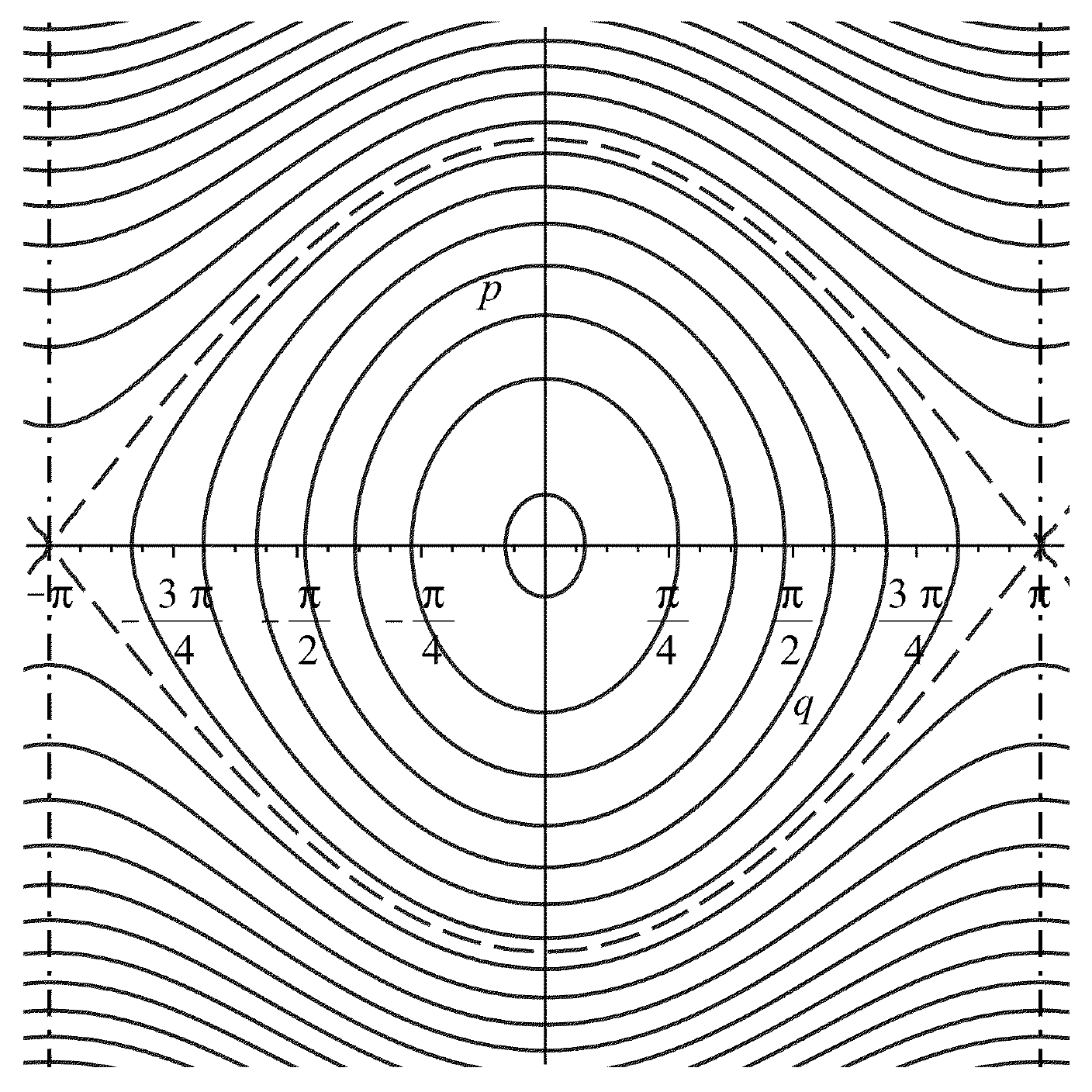}\\
\footnotesize Fig.~1: Phase portrait of the pendulum.
\end{subfigure}\hfil\begin{subfigure}[t]{0.45\textwidth}
\centering
\includegraphics[width=0.9\linewidth]{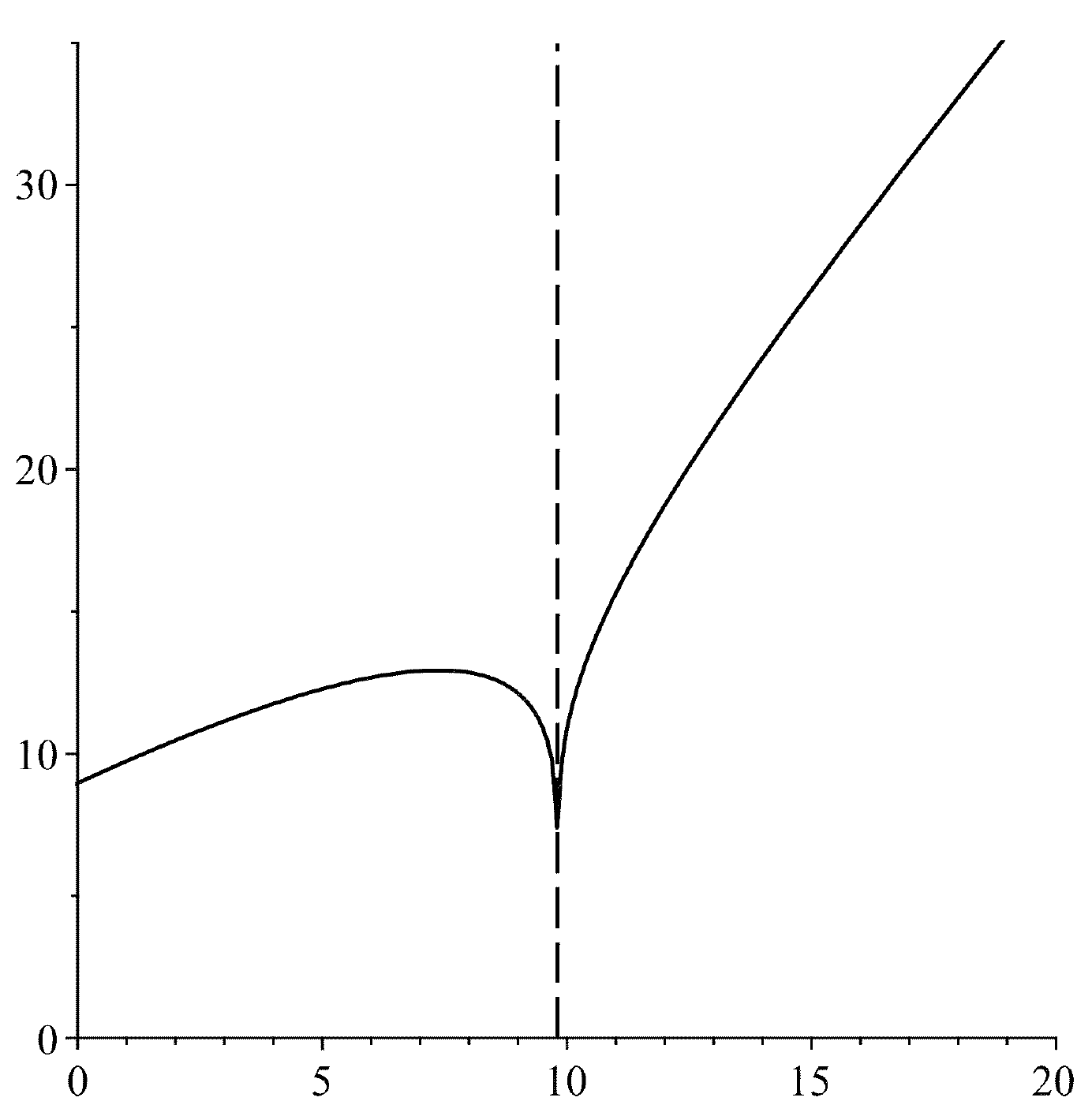}\\
\footnotesize Fig.~2: $\ave{f_{22}}=kT$ as a function of $E$. \\ The dashed line marks the critical energy $E=g$ \\(see also Fig.1.6a in \cite{Berd97}).
\end{subfigure}
\begin{subfigure}[t]{0.55\textwidth}
\centering
\includegraphics[width=0.9\linewidth]{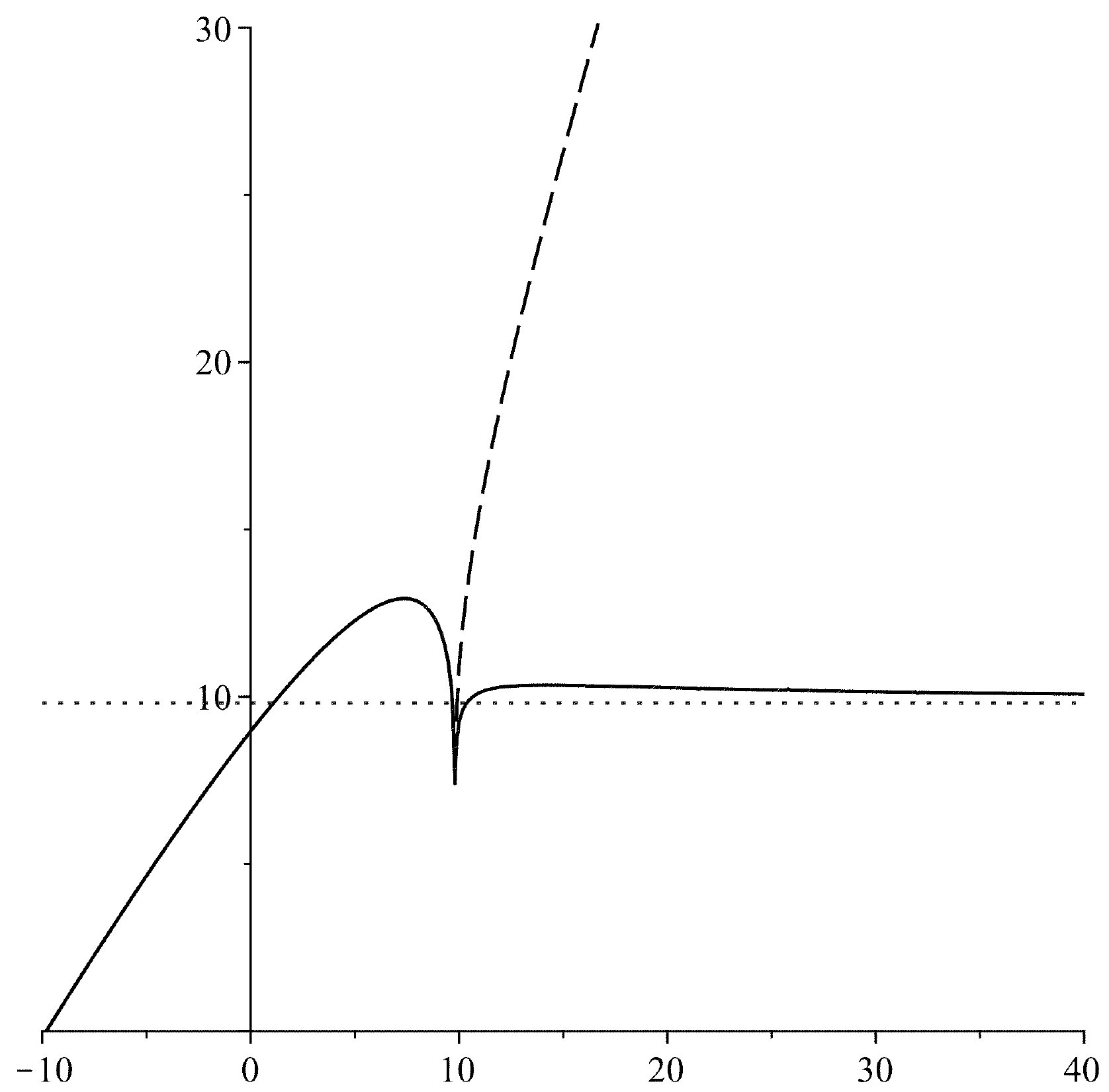}\\
\footnotesize Fig.~3: $\ave{f_{11}}$ as a function of the energy (solid line).\\ The dashed line is the value of $kT$: the two lines coincide for $E<g$.
\end{subfigure}\hfil\begin{subfigure}[t]{0.45\textwidth}
\centering
\includegraphics[width=0.9\linewidth]{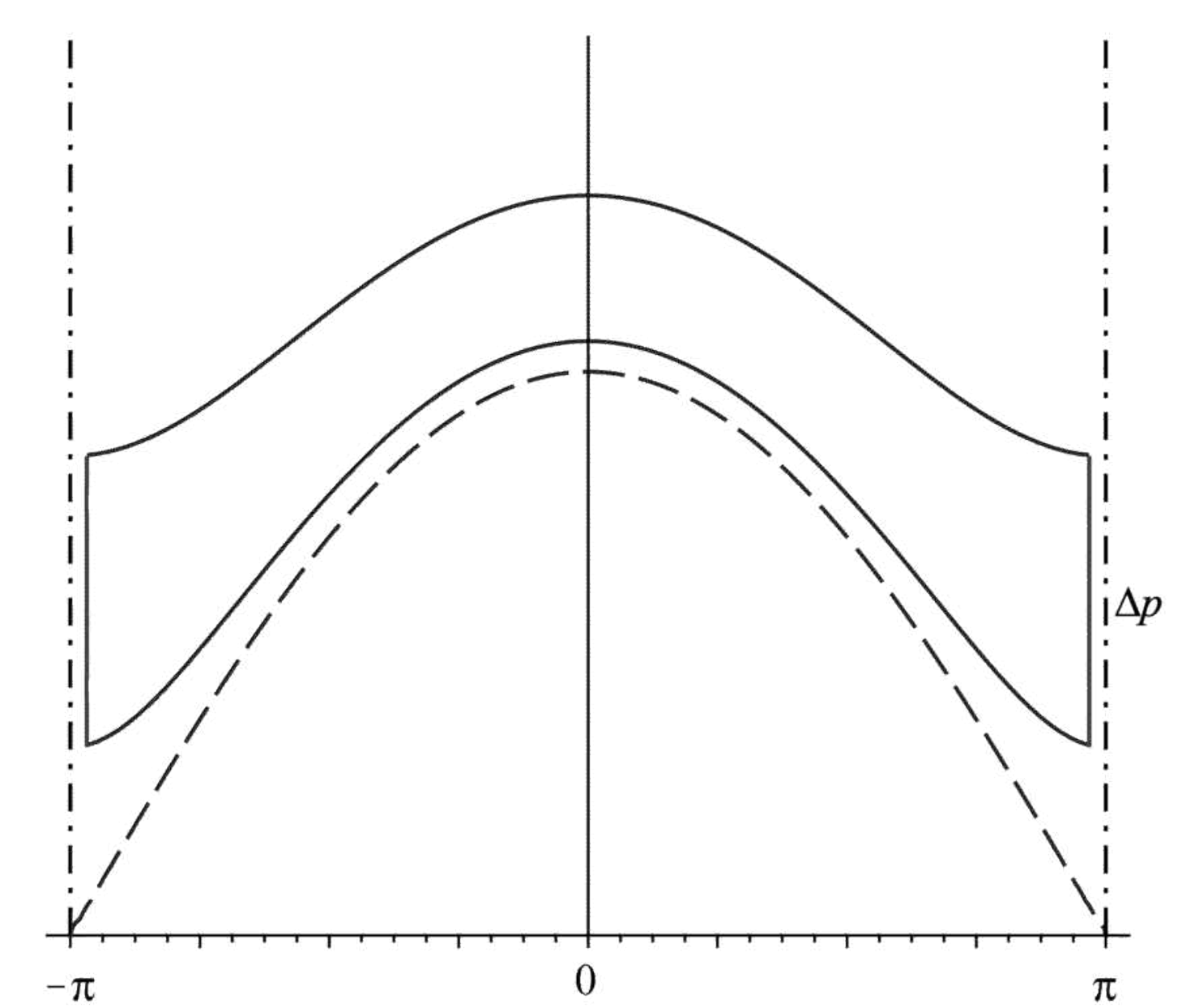}\\
\footnotesize Fig.~4: The line $\Gamma$. The dashed line is the upper branch of the separatrix.
\end{subfigure}
\end{figure}

\section*{Acknowledgements}
We are grateful to Luigi Galgani for an intense discussion about  fundamental aspects of equipartition, to Michele Caselle, Franco Magri and Lamberto Rondoni for useful advice, and to the reviewer of AoP for drawing our attention to the references \cite{Berd93,Berd97}.
\bibliographystyle{unsrt}
\bibliography{Biblio}
\end{document}